\documentclass[showpacs,preprintnumbers,amsmath,amssymb,prb]{revtex4}
\usepackage{graphicx}
\usepackage{dcolumn}
\usepackage{bm}

\def\Rey{\mbox{\it Re}}   
\def\Pran{\mbox{\it Pr}}  

\def\etal{\mbox{\it et al.\ }}

\begin{document}

\title{Small scale structure of homogeneous turbulent shear flow}

\author{D. Livescu} 
\affiliation{University of California,\\ 
Los Alamos National Laboratory, CCS-2 MS B296, Los Alamos NM 87545\\
livescu@lanl.gov}
\author{C. K. Madnia}
\affiliation{Department of Mechanical and Aerospace Engineering,\\
State University of New York at Buffalo, Buffalo, NY 14260\\
madnia@eng.buffalo.edu}

\date{April 21, 2004}

\begin{abstract}
The structure of homogeneous turbulent shear flow is studied using data
generated by Direct Numerical Simulations (DNS) and a linear analysis for both compressible
and incompressible cases. At large values of the mean shear rate, the Rapid Distortion Theory 
(RDT) limit is approached. Analytical solutions are found for the inviscid compressible RDT 
equations at long times. The RDT equations are also solved numerically for both inviscid and
viscous cases. The RDT solutions, confirmed by the DNS results, show that the even 
order transverse derivative moments of the dilatational and solenoidal velocity fields are 
anisotropic, with the dilatational motions more anisotropic than their solenoidal counterparts. 
The results obtained for the incompressible case are similar to those obtained for the solenoidal 
motions in the compressible case. The DNS results also indicate an increase in the 
anisotropy of the even order transverse derivative moments with the order of the moment, in agreement with 
the RDT predictions. Although the anisotropy decreases with Reynolds number,
it is likely that for higher even order moments it will persist at large values of the Reynolds number,
in contrast with the postulate of local isotropy. The RDT solutions also predict that the normalized
odd order transverse derivative moments of the solenoidal velocity for the compressible case and 
of the velocity for the incompressible case should approach a constant different than zero at large
times. This prediction is supported by the DNS data. For higher odd order normalized moments, the RDT 
analysis suggests that the anisotropy may persist at large values of the Reynolds number, in agreement with the 
experimental data. The amplification of the dilatational kinetic energy in the direction of the mean 
shear and the anisotropy of the dilatational dissipation tensor found in the DNS results are also consistent with 
the RDT analysis. 
\end{abstract}

\pacs{47.27.Ak,47.27.Eq,47.27.Gs,47.40.-x}

\maketitle


\section{Introduction}
\label{intro}

Most turbulent flows are anisotropic at large scales. Homogeneous shear flow represents one of the 
simplest anisotropic flows and its study can reveal important aspects of the structure and production 
of the turbulent fluctuations. The high Reynolds number experiment of Shen and Warhaft~\cite{SW00} 
indicates that the higher odd order moments of the velocity derivatives may not be consistent with the postulate
of local isotropy, which requires that the normalized odd order transverse derivative moments
approach zero at large Reynolds numbers. This finding has important consequences, since this postulate has
been central for turbulence theories and models.~\cite{SA97} Although DNS at high Reynolds numbers 
are not feasible yet, persistent anisotropy in the skewness of velocity derivatives has 
been observed earlier in the numerical results of Pumir,~\cite{Pumir96} also confirmed by 
Schumacher.~\cite{Schumacher01} 

The assumption of local isotropy also requires relations between the even order transverse derivative 
moments, which should satisfy $\left<\left(\frac{\partial u_1}{\partial x_2}\right)^{2n}\right>=
\left<\left(\frac{\partial u_1}{\partial x_3}\right)^{2n}\right>$, and similar relations for $u_2$ and 
$u_3$. There is experimental evidence that the second order
transverse derivative moments are not isotropic, especially when the mean rate of strain is 
significant.~\cite{BA87,ZALCA03} Nevertheless, the anisotropy of the higher even order transverse  
moments of the velocity derivatives has not been examined. In this study we point out a lack 
of isotropy of the even order transverse derivative moments in homogeneous shear flow for both incompressible 
and compressible cases with increased anisotropy for higher order moments. We also show that the 
small scale dilatational motions are more anisotropic than their solenoidal counterparts, so that 
the anisotropy increases for the compressible case. Moreover, the anisotropy of the even order transverse
derivative moments found in the DNS results is predicted by the RDT solutions. 
As the Reynolds number increases, the anisotropy among the transverse derivative moments
is expected to decrease. Since it is shown that the anisotropy increases with the order of the moment,
some anisotropy may persist for higher order moments at large values of the Reynolds number.
In addition, it is shown that the linear analysis predicts that the normalized odd order transverse
derivative moments of $u_1$ should approach a constant different than zero at large times, in agreement with the DNS data. 
The RDT results suggest a persistent anisotropy of the  higher odd order normalized transverse derivative moments 
at large Reynolds number, in agreement with the experimental findings.~\cite{SW00}

There are numerous experimental investigations of incompressible homogeneous shear 
flow,~\cite{TC81,TK89,GW97,SW00} however, no such study exists for the compressible case.
Compressibility effects in turbulent flows are important in many practical applications
ranging from combustion processes, to high speed aerodynamics, and to astrophysical
phenomena. Although compressible turbulence has been the subject of intense research
in the last 50 years, much less is known in comparison with incompressible
turbulence.~\citep{Lele94} In order to isolate the volume changes of the fluid elements,
the velocity field is usually decomposed into a solenoidal (divergence free) part and
a dilatational part. For homogeneous flows this decomposition is unique up to an
additive constant, which can be taken to be zero without the loss of generality.
Such decomposition in spectral space has been exploited by
Rapid Distortion Theories in studies of shock-turbulence interaction and homogeneous
turbulence subject to bulk compression or uniform mean shear.~\citep{CS99}

For the RDT equations of incompressible homogeneous shear flow, analytical solutions in wavenumber
space are known for the velocity field in both viscous and inviscid cases.~\citep{Townsend76} 
The integration of these solutions over the wavenumber space can yield predictions for various 
quantities in real space. In general, only for very early and long times the integrals
can be evaluated analytically. Long time solutions for the velocity variances and shear stress
are obtained by Moffatt~\citep{Moffatt65} and Rogers.~\citep{Rogers91} 
The early time evolution of the flow is correctly captured by the RDT 
equations since the non-normal amplification mechanism is linear in 
time.~\citep{HKW95,Waleffe97} However, due to the neglecting
of the nonlinear terms and thus the energy cascade to small scales, the RDT 
predictions can become eventually very different than those of the full 
nonlinear equations. Nevertheless, for the incompressible case, the RDT 
equations correctly capture the behavior of various correlations coefficients 
in the fully developed flow field.~\citep{Rogers91}

For the case of compressible homogeneous shear flow in the RDT limit no analytical solutions are known. 
Simone \etal~\cite{SCC97} performed RDT simulations of homogeneous shear flow and showed 
that the role of the distortion Mach number, $M_d$, on the time variation
of the turbulent kinetic energy is consistent with that found in the DNS results. They also
identified different time regimes in which the various contributions to the terms in the
RDT equations in spectral space change qualitatively, which might be the reason for the difference in the
early and long time influence of $M_d$ on the kinetic energy growth. However, in the
absence of analytical solutions, it is difficult to predict how each wavenumber will affect
the result of the integration over the wavenumber space. For example, for the incompressible case, 
Rogers~\citep{Rogers91} shows that at long times most of the contribution to the velocity variances 
and shear stress comes from a very narrow region in wavenumber space, which shrinks in time. 

The DNS results for compressible homogeneous shear flow of Blaisdell \etal~\cite{BMR91} and 
Livescu \etal~\cite{LJM02} reveal that the explicit dilatational effects tend to occur predominantly 
in the direction of the shear. Livescu and Madnia~\cite{LM01} examined the anisotropies of the 
solenoidal and dilatational motions for different values of the Reynolds and turbulent Mach numbers 
and found higher levels of anisotropy for the normal components of the dilatational dissipation 
tensor compared with those of the normal components of the solenoidal dissipation tensor. By examining the 
energy transfer leading to 
the anisotropy of the dilatational motions, Livescu \etal~\cite{LJM02} showed that the nonlinear terms 
in the transport equation for the kinetic energy components do not have a significant contribution. 
This suggests that a linear mechanism might be responsible for the amplification of the dilatational
motions in the direction of the shear.

The present study thus aims: (i) to point out a lack of isotropy of the even order transverse derivative 
moments for both compressible and incompressible cases and show that this anisotropy is due to a linear 
mechanism, (ii) to examine the ability of the linearized equations to predict the persistent anisotropy 
of the normalized odd order transverse derivative moments found in the experimental data, and
(iii) to clarify if the amplification of the dilatational kinetic energy and dilatational dissipation in 
the direction of the shear are also due to a linear mechanism. Moreover, 
the different levels of anisotropy of the small scale dilatational and solenoidal motions for
the compressible case is also discussed. Additionally, analytical solutions are presented for the 
RDT equations of compressible homogeneous turbulent shear flow.

The paper is organized as follows. Section II describes the governing equations and their
linearization in the RDT limit. The numerical methodology for solving the full nonlinear and the RDT 
equations with the parameters for the cases considered is presented in section III. Section
IV contains an analysis of the RDT equations for compressible homogeneous shear flow. The inviscid
RDT equations are solved analytically in the incompressible limit ($M_{t_0}S\rightarrow 0$) and
pressure release limit ($M_{t_0}S\rightarrow \infty$). For finite values of $M_{t_0}S$ and
large times, analytical solutions are presented for the Fourier modes of pressure and velocity
components. In section V the analytical and numerical solutions of the RDT equations are compared
with the DNS results. It is shown that the amplification of the dilatational kinetic
energy in the direction of the shear and the anisotropy of the normal components of the dilatational 
dissipation tensor are captured by the RDT equations. Also in this section, the anisotropy of the even
and odd order transverse derivative moments of the solenoidal and dilatational velocity fields for the 
compressible case and of the velocity field for the incompressible case are studied using DNS results 
and shown to be predicted by the RDT analysis. Summary and conclusions are given in section VI.

\section{Governing equations }
\label{goveq}

The conservation equations which provide the mathematical model of the
problem are the continuity, Navier-Stokes, and energy transport equations. For
a compressible homogeneous shear flow (for which the mean, Favre averaged, velocity is given by
$\tilde{u}_1=S x_2$ with $S$ constant, $\tilde{u}_2=0$, and $\tilde{u}_3=0$)
after Rogallo's transformation of coordinates~\citep{Rogallo81} 
$x_i'=B_{ij}x_j$, these equations, in their non-dimensional form, 
become~\citep{BMR93,LJM02}

\begin{equation}
\rho_{,t}+(\rho u_j'')_{,k} B_{kj} =0
\label{coeq1}
\end{equation}

\begin{eqnarray}
(\rho u_i'')_{,t}=&-&\rho u_2'' S \delta_{i1}-(\rho u_i'' u_j'')_{,k}B_{kj}
  -p_{,k} B_{ki}\nonumber \protect\\
&+&[(\tau_{ij}'+\frac{\mu}{Re_0}(S\delta_{i1}\delta_{j2}+
  S\delta_{i2}\delta_{j1}))]_{,k}B_{kj}
\label{moeq1}
\end{eqnarray}

\begin{eqnarray}
(\rho \phi)_{,t}&=& S(\tau_{12}'-\rho u_1''u_2'')-(\rho u_j'' \phi)_{,k} B_kj
 -(p u_j'')_{,k} B_{kj}\nonumber \protect\\
 &+&(\tau_{ij}'u_i'')_{,k} B_{kj}+\frac{1}{(\gamma -1)
   M_0^2 Re_0 Pr}(\mu T_{,l}B_{lj})_{,k}B_{kj}.
\label{eneq1}
\end{eqnarray}

\noindent
The stress tensor is $\tau_{ij}'=\frac{2\mu}{Re_0}(s_{ij}'-
\frac{1}{3}\Delta'\delta_{ij})$, where $s_{ij}'=\frac{1}{2}(u_{i,k}''B_{kj}+
u_{j,k}''B_{ki})$ is the strain rate tensor, and $\Delta'=u_{i,k}''B_{ki}$ is 
the dilatation of the velocity fluctuations. The primary transport variables 
are the density $\rho$, velocity fluctuations (with respect to Favre average) in 
$x_i$ direction $u_i''$, and modified total energy $\phi\equiv \frac{p}{\rho(\gamma-1)}+
\frac{1}{2}u_i''u_i''$, where $p$ is the instantaneous pressure and the ratio 
of the specific heats $\gamma=1.4$. The coordinate transformation matrix has 
constant diagonal components $B_{ii}=\beta_i$ and the only off-diagonal nonzero
component is $B_{12}=-\beta_1 St$. 

The thermodynamic variables are related through the equation of state,
$p=\frac{\rho T}{\gamma M_0^2}$, and the nondimensional viscosity is
modeled by assuming a power law $\mu=T^n$, with $n=0.7$. The reference scales 
used to nondimensionalize the governing equations are the initial rms velocity fluctuations
($u_0=\sqrt{u_{i0}''u_{i0}''}$), initial mean temperature ($T_0$),
initial mean density ($\rho_0$) and a reference length-scale ($l_0$).
Consequently, the non-dimensional parameters in Eqs. (\ref{coeq1})-
(\ref{eneq1}) are the computational Reynolds number, $\Rey_0=\frac{\rho_0 u_0 l_0}{\mu_0}$, 
the Prandtl number, $\Pran=\frac{\mu_0 c_p}{\kappa_0}$, and the reference Mach number,
$M_0=\frac{u_0}{\sqrt{\gamma R T_0}}$, where $R$ is the gas constant and $c_p$ is the 
specific heat at constant pressure. The reference viscosity, $\mu_0$, and thermal diffusivity, 
$\kappa_0$, are assumed to be proportional to $T_0^n$. 

For the incompressible case, the energy equation is eliminated, the continuity equation is
replaced by the incompressibility constraint, and the molecular viscosity is constant. The 
governing equations become

\begin{equation}
u'_{j,k}B_{kj}=0
\label{ieq1}
\end{equation}

\begin{equation}
u'_{i,t}=-u'_2S\delta_{i1}-(u'_iu'_j)_{,k}B_{kj}-p_{,k}B_{ki}+\tau'_{ij,k}B_{kj},
\label{ieq2}
\end{equation}

\noindent
where $u'_i$ are the velocity fluctuations and 
$\tau'_{ij}=\frac{1}{Re_0}(u'_{i,k}B_{kj}+u'_{j,k}B_{ki})$.

\subsection{Linearized equations for compressible homogeneous turbulent shear flow}
\label{goveqlin}

The linearized nondimensional equations of motion for density, $\rho'$, velocity, $u_i''$, 
and pressure, $p'$, fluctuations in a compressible homogeneous turbulent flow are derived
by Kovasznay.~\cite{Kovasznay53} For the case of a homogeneous turbulent shear flow these 
equations correspond to the RDT limit, which assumes large values for the mean 
shear rate $S$. In the moving coordinate system defined above,
the linearized equations become

\begin{eqnarray}
\rho_{,t}'&=&-<\rho>\Delta'
\label{colin}\\
u_{i,t}''&=&-\frac{1}{<\rho>}p_{,k}'B_{ki}-Su_2'' \delta_{i1}+
 \frac{2 <\mu>}{<\rho> Re_0}(s_{ij}'-\frac{1}{3}\Delta' \delta_{ij})_{,k}B_{kj}
\label{molin}\\
p_{,t}'&=&-\gamma <p> \Delta'+\frac{4<\mu>(\gamma-1)}{Re_0}S s_{12}'
  \nonumber \protect \\
 &&+\frac{<\mu>}{M_0^2 Re_0 Pr>}<T>\left(\frac{p'_{,jj}}{<p>}-
    \frac{\rho'_{,jj}}{<\rho>}\right).
\label{enlin}
\end{eqnarray}

While the average density is constant in time for homogeneous shear flow, the 
linearized equation for the mean pressure is

\begin{equation}
<p>_{,t}=\frac{<\mu>(\gamma-1)}{Re_0}S^2,
\label{meanp}
\end{equation}
with the right hand side of the equation given by viscous dissipation of the
mean flow.

Although in general the linearized equations form a coupled system of equations, for the inviscid 
case they can be decoupled (note that in this case the mean pressure becomes constant). 
Thus, an equation for the dilatation can be obtained by taking the divergence of Eq. (\ref{molin})

\begin{equation}
\Delta_{,t}'=-\frac{1}{<\rho>}p_{,jk}'B_{ji}B_{ki}-2Su_{2,k}''B_{k1}.
\label{lindil}
\end{equation}

\noindent
This equation is used to eliminate the dilatation from Eq. (\ref{enlin}). 
Finally, after eliminating $u_2''$ with the use of Eq. (\ref{molin}), an 
equation for the pressure fluctuations is obtained

\begin{equation}
\frac{p'_{,ttt}}{c_0^2}=p'_{,tjk}B_{ji}B_{ki}-4Sp'_{,jk}B_{j1}
 B_{k2},
\label{p1}
\end{equation}

\noindent
where $c_0=(\gamma <p>/<\rho>)^{1/2}$ (which for the inviscid
case becomes constant). Equation (\ref{p1}) is a third order linear differential equation in time, with 
variable coefficients. For $S=0$ (decaying isotropic turbulence) Eq. 
(\ref{p1}) leads to a wave-like equation, with a bounded solution. For 
$S\neq 0$, the time $t$ can be nondimensionalized by $1/S$, so that $\eta \equiv St$.
Furthermore, after applying the Fourier transform, Eq. (\ref{p1}) yields

\begin{equation}
\hat{p}_{,\eta\eta\eta}=-\frac{c_0^2}{S^2}(k^2 \hat{p}_{,\eta}-4 k_1 k_2\hat{p}),
\label{p2}
\end{equation}

\noindent
where the wavenumber $k_j$ is related to the wavenumber $\hat{k}_i$ in
the moving coordinate system by $k_j=B_{ij}\hat{k}_i$, and $k^2=k_jk_j$. The 
notations used in the present paper for the wavenumbers in the fixed and moving
coordinate systems are the opposite of the notations used by
Rogers~\citep{Rogers91}. Due to the time variation of $k_2$, Eq. (\ref{p2})
does not have an easily derived analytical solution except in the case
$k_1=0$ and for very early and long times ($\eta \approx 0$ and $\eta \rightarrow \infty$).
Nevertheless, it should be noted that, although for the full nonlinear equations 
the initial turbulent Mach number, $M_{t_0}=(2 K_0)^{1/2}/c_0$, and $S$ are independent 
parameters, they appear together in Eq. (\ref{p2}), since 
$c_0^2/S^2=2 K_{0}/<\rho>/(M_{t_0} S)^2=1/(M_{t_0} S)^2$ (note that the
initial value of the turbulent kinetic energy is $K_{0}=0.5$ and
$<\rho>=1$). Moreover, if the distortion Mach number,~\cite{SCC97} $M_d=S L/c$, 
is introduced, then $c_0^2/S^2=L_0^2/M_{d_0}^2$, where $M_{d_0}$ is the initial
value of the distortion Mach number and $L_0$ is the initial value of the integral
scale $L=(2 K)^{3/2}/\epsilon$. Similar equations, depending on $c_0^2/S^2$, $\eta$, the wavenumber,
and initial conditions can be found for the Fourier modes of dilatation, density
fluctuations, and velocity components. In order to complete the
analysis, the initial conditions for Eq. (\ref{p2}) are considered.
These relations can be rearranged to isolate the parameter $(M_{t_0}S)$ by
making the transformation $\hat{p}'=M_{t_0} \hat{p}$,

\begin{eqnarray}
\hat{p}'|_{\eta=0}&=&M_{t_0}\hat{p}_0
\label{p1in}\\
\hat{p}'_{,\eta}|_{\eta=0}&=&-\frac{2K_0}{<\rho>M_{t_0} S}\hat{d}_0
\label{p2in}\\
\hat{p}'_{,\eta\eta}|_{\eta=0}&=&\left(-\frac{k'^2}{(M_{t_0}S)^2}
 (M_{t_0}\hat{p}_0)+\frac{2 k_1'}{M_{t_0} S} \mbox{\boldmath{$i$}} 
  \hat{u}_{2_0}\right)\frac{2K_0}{<\rho>},
\label{p3in}
\end{eqnarray}

\noindent
where $k_i'$ is the initial value of the wave number in the fixed
coordinate system ($k_i'=\hat{k_i}$), $k'^2=k_i'k_i'$, 
$\mbox{\boldmath{$i$}}=\sqrt{-1}$, and $\hat{d}_0$ and $\hat{u}_{2_0}$ are 
the initial values of the Fourier modes of the
dilatation and velocity fluctuations in $x_2$ direction, respectively.
It can be seen that, if the initial value of the pressure mode scales
with $1/M_{t_0}$ for constant $(M_{t_0}S)$, then $\hat{p}'$
depends on $(M_{t_0}S)$ and $\eta$ only. This condition is fulfilled if the 
pressure fluctuations are initialized with zero or from a Poisson equation, which 
is the case with most DNS studies of homogeneous turbulent shear flow. Note that
to first order the Poisson equation becomes

\begin{equation}
M_{t_0}\hat{p}=2 \mbox{\boldmath{$i$}} <\rho> (M_{t_0}S)\frac{k_{1}}{k^2}\hat{u}_{2}.
\label{poisson}
\end{equation}

\noindent
In this case, from the inviscid momentum equations and Eq.~\ref{lindil} 
for the dilatation it yields, after replacing $t$ with $\eta$, that $\hat{u}_i$ and
$\hat{\Delta}$ depend only on $(M_{t_0}S)$. Moreover, if the initial values of the 
density fluctuations are set to zero, $S \hat{\rho}$ will also be a function of 
$(M_{t_0}S)$, wavenumber, and $\eta$ only.

As a result, for the initializations usually used in the DNS of homogeneous 
turbulent shear flow, the inviscid linearized theory predicts the existence of only
one compressibility parameter, $(M_{t_0}S)$, for the properly scaled variables. 
However, the addition of either the nonlinear or the viscous terms prevents the 
governing equations to be written in terms of only one compressibility parameter. 
Therefore, for moderate values of $S$ or Reynolds number, two independent 
compressibility parameters can be defined. Nevertheless, as pointed out by Simone 
\etal,~\cite{SCC97} changes in $M_{d_0}$ at constant $M_{t_0}$ are more important 
to the growth of turbulent kinetic energy than changes in $M_{t_0}$ at constant 
$M_{d_0}$.

\section{Numerical solution procedure}
\label{linnum}

\subsection{DNS methodology}

Equations (\ref{coeq1})-(\ref{eneq1}) are solved using the Fourier pseudo-spectral
method with Rogallo's remeshing technique.~\citep{Rogallo81} Details can be 
found in Livescu \etal~\cite{LJM02} All simulations are performed 
within a box containing $256^3$ points. The computational domain is twice as long 
in the stream-wise direction as in the cross-stream and span-wise directions so 
that $\beta_1=0.5$, $\beta_2=1.0$ and $\beta_3=1.0$. The initial velocity is a 
random solenoidal field with unity r.m.s. and the spectrum given by 
$E(k)=k^4/k_{0v}^5\ exp(-2k^2/k_{0v}^2)$ with $k_{0v}=10$. The initial pressure 
fluctuations are evaluated from a Poisson equation, the initial density field is uniform, 
and the initial value of the mean pressure is calculated from the mean equation of state.
The numerical method, mesh size and initial conditions for pressure and velocity fields
are the same for the incompressible case.

Several cases, 1 to 5, are considered at different values of the mean shear 
rate. For each case two simulations, one compressible and one incompressible, are performed. 
Table I presents the relevant information 
for each of the cases considered. The initial nondimensional shear rate is defined as $S_0^*=2KS/\epsilon$. 
For the compressible cases, the distortion Mach number $M_{d_0}$ is also listed. Cases 1 to 5 have the
initial value of the Taylor Reynolds number $Re_{\lambda_0}=21$ and it
increases to $Re_\lambda \sim 100$ at the end of the simulations.
Additionally, one more case (case 6) is considered with $Re_{\lambda_0}=50$, 
and all the other parameters as those of case 2. In the next sections, 
the compressible simulations corresponding to cases 1 to 6
are labeled C1 to C6, and the incompressible cases are labeled I1 to I6.
All simulations were monitored to ensure that the 
integral scales remain small compared to the box size and the Kolmogorov 
microscale is larger than the grid size. The compressible runs were also repeated with
initial turbulent Mach numbers ranging from $M_{t_0}=0.1$ to $M_{t_0}=0.6$ and all the
results presented in the next sections remained qualitatively unchanged.

Based on the magnitude of the shear rate, a simple scaling argument can be used to
characterize the relative importance of the nonlinear term in the momentum equations 
compared to the linear term arising from the presence of the nonzero mean velocity
gradient. Thus, as shown in Ref.~\onlinecite{SRY03}, the nonlinear term is negligible
when $S^*_1/Re_{\lambda_1}\gg 1/8$ and dominates when $ S^*_1/Re_{\lambda_1}\ll 0.35$,
where $S^*_1=u'^2S/\epsilon$ and $Re_{\lambda_1}=u'^2/(\nu\sqrt{<u_{1,1}^2>})$,
with $u'=\sqrt{<u_1^2>}$. Figure 1 shows the time evolution of 
$S^*_1/Re_{\lambda_1}$
for the compressible cases considered. The results obtained for the incompressible
cases are close. The initial range of variation covers the region in which both
the linear and nonlinear terms in the momentum equations are important, with case C1, 
with lowest shear, for which the nonlinear effects dominate and case C5, with highest
shear, for which the linear term becomes more important. After a development time 
the results obtained for the cases with different values of the mean shear rate become
close, while retaining the dependence on the initial Reynolds number. For all cases, 
the late time values of $S^*_1/Re_{\lambda_1}$ are in the region where the nonlinear
effects dominate, based on the scaling arguments given in Ref.~\onlinecite{SRY03}.

\subsection{RDT methodology}

The unknowns in Eqs. (\ref{colin})-(\ref{enlin}) are transformed into~\cite{SCC97}

\begin{eqnarray}
\hat{f}_1(\mbox{\boldmath{$k$}},\eta)&=&\frac{k_1\hat{u}_3-k_3\hat{u}_1}{m}\\
\hat{f}_2(\mbox{\boldmath{$k$}},\eta)&=&(k_1\hat{u}_1+k_3\hat{u}_3)\frac{k_2}
 {mk}-\frac{m}{k}\hat{u}_2\\
\hat{f}_3(\mbox{\boldmath{$k$}},\eta)&=&\frac{k_1\hat{u}_1+k_2\hat{u}_2+
  k_3\hat{u}_3}{k}=\frac{\hat{\Delta}}{\mbox{\boldmath{$i$}}k}\\
\hat{f}_4(\mbox{\boldmath{$k$}},\eta)&=&\mbox{\boldmath{$i$}}\hat{p}\\
\hat{f}_5(\mbox{\boldmath{$k$}},\eta)&=&\mbox{\boldmath{$i$}}\hat{\rho},
\end{eqnarray}

\noindent
where $m=\sqrt{k_1^2+k_3^2}$.
$\hat{f}_1$ is the component of the velocity along the perpendicular
on the plane defined by the wavenumber vector and $k_2$ direction.
$\hat{f}_2$ is the component along $k_2$ direction so that $\hat{f}_1$
and $\hat{f}_2$ form the solenoidal velocity and $\hat{f}_3$ is the
projection of the velocity on the wavenumber vector and corresponds to
the dilatational part of the velocity. Additionally, since the
equations are linear, a matrix exponentiation method can be used by
letting $\hat{f}_i(\mbox{\boldmath{$k$}},\eta)=g_{ij}
(\mbox{\boldmath{$k$}},\eta)\hat{f}_j(\mbox{\boldmath{$k$}},0)$, with
$g_{ij}(\mbox{\boldmath{$k$}},0)=\delta_{ij}$.

The resulting system for the unknowns $g_{ij}$ becomes
\begin{eqnarray}
g_{1j,\eta}&=&-\frac{k_3}{k}g_{2j}+\frac{k_2k_3}{mk}g_{3j}-
  \frac{<\mu>}{Re_0<\rho>S}k^2g_{1j}
\label{g1}\\
g_{2j,\eta}&=&\frac{k_1k_2}{k^2}g_{2j}-\frac{k_1}{m}g_{3j}-
  \frac{<\mu>}{Re_0<\rho>S}k^2g_{2j}\\
g_{3j,\eta}&=&2\frac{k_1m}{k^2}g_{2j}-\frac{k_1k_2}{kk^2}g_{3j}-
  \frac{k}{<\rho>S}-\frac{4}{3}\frac{<\mu>}{Re_0<\rho>S}k^2g_{3j}\\
g_{4j,\eta}&=&\frac{\gamma<p>}{S}k g_{3j}+\frac{2<\mu>}{Re_0}(\gamma -1)
  \left(-\frac{k_1k_3}{m}g_{1j}+\frac{k_1k_2^2}{km}g_{2j}+
  2\frac{k_1k_3}{k}g_{3j}\right)\protect \nonumber\\
 &&-\frac{<\mu>}{M_{t_0}^2 Re_0 Pr S}k^2\left(\frac{g_{4j}}{<p>}-
   \frac{g_{5j}}{<\rho>}\right)\\
g_{5j,\eta}&=&\frac{<\rho>}{S}kg_{3j},
\label{g5}
\end{eqnarray}

\noindent
where $j=\overline{1,5}$. The system \ref{g1}-\ref{g5} is transformed
into spherical coordinates by setting $k_1'=k'\cos \alpha$,
$k_2'=k'\sin \alpha \sin \phi$, and $k_3'=k'\sin \alpha \cos \phi$.
Cases with $(101\times 602\times 204)$ grid points for the
coordinates $(k',\alpha,\phi)$ are considered. The discretization is
nonuniform, allowing for more grid points near the line
$\alpha=\phi=\pi/2$. This is necessary, as explained in the next
section, in order to correctly capture a local peak of some of the 
unknowns, which gives an important contribution to
their integral over the wavenumber space. A fourth order Runge-Kutta
scheme is employed for the time advancement. The equations are also
solved for the incompressible case (which is recovered by letting
$\hat{f}_3=0$). It should be noted that
the system \ref{g1}-\ref{g5} is independent of the initial spectra, so
that it is solved only once for a particular set of parameters.
Furthermore, when integrating over the wavenumber space only the
initial power spectra are required, and not the phase information.

The linearized Eqs. (\ref{g1})-(\ref{g5}) are solved for different values of 
$S_0^*$ ranging from $7.24$ to $362$. For all compressible cases $M_{t_0}=0.3$, 
so that the range in $S_0^*$ corresponds to $2.2<M_{d_0}<109$. This range extends from a 
low shear case in which the 
nonlinear effects in the corresponding DNS case are important to cases with very 
high shear. The initial velocity power spectrum is the same as in the DNS simulations 
described earlier. For the compressible case, for each value of $(M_{t_0}S)$ four 
types of initial conditions, labeled L1 to L4, are considered (Table II). For cases L1 and L3 the 
pressure fluctuations are initialized from a Poisson equation, while for cases 
with L2 initial conditions they are set to zero. The initial velocity field is 
solenoidal for cases L1 and L2 while $\chi_0=K_{d0}/K_{0}=0.1$ for cases L3
and L4, where $K_{d0}$ is the initial dilatational kinetic energy. For cases 
with L4 initial conditions the strong form of the acoustic equilibrium is 
considered,~\citep{SCC97} so that there is an equipartition of energy between the
potential and kinetic energies of the dilatational component at each wavenumber. 
Therefore the initial pressure power spectrum is given by 
$E_p(k')=\sqrt{\gamma <p>|_{t=0}}\ <\rho_0>E_d(k')$, where $E_d$ is the initial power 
spectrum of the dilatational velocity. For each case, both the inviscid
and viscous equations are solved. The types of initial conditions considered are
not exhaustive for the initial conditions used in DNS of compressible
turbulence (e.g. see Refs.~\citep{RB97,SPK01} for decaying isotropic turbulence).
However, they were chosen such that the role of a nonzero initial dilatational 
field can be isolated. In the next sections, the numerical solutions of the RDT 
equations are compared with the DNS results and with the analytical predictions.

\section{Analytical results}
\label{asymptot}

By examining Eq. (\ref{p2}) for the pressure mode
three cases can be identified based on the values of $1/(M_{t_0}S)^2$.

\subsection{``Pressure-released'' limit ($\frac{1}{(M_{t_0}S)^2} \rightarrow 0$)} 

This case corresponds to values of the distortion Mach number
approaching infinity. After integrating Eq. (\ref{p2}), a quadratic
dependence on $\eta$ is obtained. However, when the initial conditions
(\ref{p1in})-(\ref{p3in}) simplified for large $(M_{t_0}S)$ are applied,
the solution is obtained as $\hat{p}=\hat{p}|_{\eta=0}$. Since the
pressure mode is multiplied by the factor $1/(M_{t_0}S)$ in the
transport equations for the velocity modes, it can be dropped from
these equations. Therefore the equations for the velocity modes become:

\begin{equation}
\hat{u}_{1,\eta}=-\hat{u}_2\mbox{,\hskip 1cm}\hat{u}_{2,\eta}=0\mbox{,\hskip 1cm}
\hat{u}_{3,\eta}=0
\end{equation}

\noindent 
with the solutions $\hat{u}_1=\hat{u}_2|_{\eta=0}\eta+\hat{u}_1|_{\eta=0}$,
$\hat{u}_2=\hat{u}_2|_{\eta=0}$, and $\hat{u}_3=\hat{u}_3|_{\eta=0}$.
After integrating over the wavenumber space, it is obtained that the kinetic energy
in $x_1$ direction increases quadratically, while it remains constant in the other 
two directions. This case corresponds to the ``pressure-released'' limit discussed 
by Cambon \etal~\cite{CMM93} and Simone \etal~\cite{SCC97}. 

\subsection{Incompressible case ($\frac{1}{(M_{t_0}S)^2} \rightarrow \infty$)} 

Since $S$ is assumed large in the RDT limit, as $\frac{1}{(M_{t_0}S)^2} \rightarrow \infty$
it yields $M_{t_0}\rightarrow 0$, so that the incompressible case is recovered. The transport
equation for the pressure mode becomes

\begin{equation}
k^2\hat{p}_{,\eta}-4 k_1 k_2 \hat{p}=0.
\label{pinc}
\end{equation}

\noindent
The solution is straightforward and can be written as

\begin{equation}
\hat{p}=\hat{p}|_{\eta=0} \frac{k'^4}{k^4},
\label{pincsol}
\end{equation}

\noindent
with $\hat{p}|_{\eta=0}=2 \mbox{\boldmath{$i$}}S\frac{k_1'}{k'^2}\hat{u}_{2}
|_{\eta=0}$. In order to find the pressure variance from Eq. (\ref{pincsol}), the 
integration over the wavenumber space should be performed numerically. Analytical 
expressions can be obtained to approximate the time behavior of the pressure 
variance only for $\eta \rightarrow 0$ (early time) and $\eta \rightarrow \infty$ 
(long time). In both cases the integration is conveniently done in spherical 
coordinates. The procedure is similar to that used by Townsend~\cite{Townsend76} 
and Rogers~\cite{Rogers91} to find the time dependence of velocity and scalar one 
point statistics in incompressible turbulent shear flow. After the transformation 
$k_1'=k'\cos \alpha$, $k_2'=k'\sin \alpha \sin \phi$, $k_3'=k'\sin \alpha \cos 
\phi$, and $d\mbox{\boldmath $k'$}=k'^2\sin \alpha\ d\alpha\ d\phi\ d k'$, for 
isotropic initial conditions (so that $|\hat{u}_2|_{\eta=0}|^2=\frac{E(k')}{4\pi
k'^2}\frac{k_1'^2+k_3'^2}{k'^2}$ with $E(k')$ the initial velocity power spectrum),
the pressure spectrum function becomes

\begin{equation}
\mbox{\boldmath $\Phi_p$}(\mbox{\boldmath $k'$})=\frac{E(k')}{4\pi k'^2}
  \frac{4 S^2\cos^2\alpha (\cos^2\alpha+\sin^2\alpha \cos^2\phi)}
  {k'^2(1-2\eta\cos \alpha \sin \alpha \sin \phi +\eta^2 \cos^2 \alpha)^4},
\label{pspec}
\end{equation}

\noindent
and the pressure variance can be found from
\begin{equation}
<p'^2>(\eta)=\frac{(2\pi)^3}{V}\int_{-\infty}^{\infty}\mbox{\boldmath$\Phi_p$}dk_i'
.
\label{pintegr}
\end{equation}

For small values of $\eta$ (early time) the integrand can be expanded in a Taylor
series in $\eta$, which can be integrated analytically. It yields

\begin{equation}
\frac{<p'^2>}{q}=\frac{4}{15}+\frac{8}{105}\eta^2+...,
\end{equation}

\noindent
where $q=4S^2 \int \frac{E(k')}{k'^2}dk'$. In terms of the initial
value of the pressure variance the dependence is
$\frac{<p'^2>}{<p_0'^2>}=1+\frac{2}{7}\eta^2+...$

As explained by Rogers,~\cite{Rogers91} the long time behavior can not be obtained 
using the same procedure, since expanding around $\eta=\infty$ yields a 
result which is not uniformly valid over the entire domain. This can be easily
seen by evaluating $\hat{p}$ for $k_2'=1$, $k_3'=0$, and
$k_1'=1/\eta$, which for $\eta \rightarrow \infty$ yields

\begin{equation}
\lim_{\eta \rightarrow \infty} \hat{p}(k_1'=1/\eta,\eta)=\infty.
\end{equation}

\noindent
Therefore a significant contribution to the integral of the pressure
spectrum over the wavenumber space comes from a narrow region near
$k_1'=0$ which decreases in size as $\eta$ increases. Following
Rogers,~\cite{Rogers91} the asymptotic behavior can be found by dividing the
domain into two regions, an inner region located around
$\eta k_1'=O(1)$ (which corresponds to $\alpha=\pi/2$) and an outer
region consisting of the rest of the domain covered by
$0\leq \alpha\leq \pi$ and $0\leq \phi \leq 2 \pi$. In the inner
region,  around $\alpha =\pi/2$ and $\phi=\pi/2$, $\mbox{\boldmath $\Phi_p$}
(\mbox{\boldmath $k'$}) k'^2\sin \alpha$ has a peak. The magnitude of the peak 
and the extent in $\alpha$ and $\phi$ over which it occurs can be found by
writing $\alpha=\pi/2-\epsilon_\alpha$ and $\phi=\pi/2-\epsilon_\phi$ and
expanding the $sin$ and $cos$ functions in the expression for the pressure 
spectrum. It yields that $\mbox{\boldmath $\Phi_p$}(\mbox{\boldmath $k'$})
k'^2\sin \alpha$ is of order $\eta^4$ for $\epsilon_\alpha=1/\eta\pm O(1/\eta^2)$
and $\epsilon_\phi=O(1/\eta)$, so that the peak occurs over a region of extent 
$O(1/\eta^2)$ in $\alpha$ and $O(1/\eta)$ in $\phi$. This indicates a linear 
increase for the pressure variance. The correct behavior can be obtained by 
combining the results from the inner and outer regions. The result obtained remains
$<p'^2>/<p_0'^2> \sim \eta$, also supported by the results of the
numerical integration.

\subsection{$\frac{1}{(M_{t_0}S)^2}$ finite} 

In this case the transport equation for the pressure mode (Eq. \ref{p2}) 
remains third order, and does not have an easily derived analytical solution. 
However, some limiting cases can be found. First, it should be observed that
$\hat{p}$ approaches zero as $k_i'\rightarrow \infty$ so that only
finite values of $k_i'$ are of interest. Then for small values of
$\eta$ the equation reduces to an equation with constant coefficients
which can be easily solved. However, since the roots of the
characteristic polynomial are dependent on the parameter $(M_{t_0}S)$,
the analytical computations involving the integration of the solution over the
wavenumber space are cumbersome. Nevertheless, the solution was
verified numerically and it agrees reasonably well with the full numerical
solution of the RDT equations.

As $\eta \rightarrow \infty$, unlike in the incompressible case, the behavior of 
the Fourier modes in the outer region ($k_1'\eta >>1$) is very important
for the integration over the wavenumber space. It is shown above that
for the incompressible case the contribution from the inner region yields 
a linear increase for the pressure variance. However, as the parameter 
$\frac{1}{(M_{t_0}S)^2}$ decreases, the peak of the pressure mode occurring 
in the inner region decreases its magnitude (Fig. 2) and the 
contribution from this region to the pressure variance yields a less than 
linear increase in time. This contribution to the pressure variance is 
negligible at long times since it is shown below that the contribution from 
the outer region leads to a linear increase in the pressure variance.

In the outer region $k_1'\eta \rightarrow \infty$ for
$\eta \rightarrow \infty$ and Eq. (\ref{p2}) becomes

\begin{equation}
\hat{p}_{,\eta\eta\eta}=-\frac{2 K_0 k_1'^2}{<\rho>(M_{t_0}S)^2}(\eta^2\hat{p}_{,\eta}+
 4\eta\hat{p}).
\label{p3}
\end{equation}

The solution of Eq. (\ref{p3}) can be written as
$\hat{p}=\phi_1\ \hat{p}|_{\eta=0}+\phi_2\ \hat{p}_{,\eta}|_{\eta=0}+
\phi_3\ \hat{p}_{,\eta\eta}|_{\eta=0}$, with $\phi_i$ given by the formulas

\begin{eqnarray}
\phi_1&=&\mbox{}_1F_2\left([1],\left[\frac{1}{2},\frac{3}{4}\right],
 -\frac{1}{4}\zeta^2\right) \\
\phi_2&=&J_{-\frac{1}{4}}(\zeta)(\zeta/2)^{3/4}2\Gamma\left(\frac{3}{4}\right)
 \left(\frac{M_{t_0} S}{k_1'}\right)^{1/2}\left(\frac{<\rho>}{2K_0}\right)^{1/4}\\
\phi_3&=&J_{\frac{1}{4}}(\zeta)(\zeta/2)^{3/4} \Gamma\left(\frac{1}{4}\right)
 \frac{M_{t_0}S}{2 k_1'}\left(\frac{<\rho>}{2K_0}\right)^{1/2},
\end{eqnarray}

\noindent
where $_1F_2$ and $J_\nu$ represent a generalized hypergeometric
function and the Bessel function of the first kind of order $\nu$,
respectively, $\Gamma$ is the gamma function and
$\zeta=\frac{k_1'\eta^2}{M_{t_0}S}\sqrt{\frac{K_0}{2<\rho>}}$. For large values of 
the nondimensional time $\eta$, using the asymptotic expansions for these functions
given in Ref.~\onlinecite{GR00}, it is obtained that

\begin{eqnarray}
\phi_1&\approx& -(\zeta/2)^{1/4}\ sin(\zeta-\frac{3 \pi}{8})
  \Gamma\left(\frac{3}{4}\right)
\label{s1}\\
\phi_2&\approx& (\zeta/2)^{1/4}\ cos(\zeta-\frac{\pi}{8})
  2 \Gamma\left(\frac{3}{4}\right)\left(\frac{M_{t_0}S}{\pi k_1'}\right)^{1/2}
  \left(\frac{<\rho>}{2K_0}\right)^{1/4}\\
\phi_3&\approx& (\zeta/2)^{1/4}\ cos(\zeta-\frac{3 \pi}{8})
  \Gamma\left(\frac{1}{4}\right)\frac{M_{t_0}S}{2 \sqrt{\pi}k_1'}
  \left(\frac{<\rho>}{2K_0}\right)^{1/2}.
\label{s3}
\end{eqnarray}

\noindent
These expressions indicate that the Fourier mode of the pressure oscillates
in time and the amplitude of the oscillations increases as $\sqrt{\eta}$. 
In order to find the pressure variance, the integration over the wavenumber space 
can be performed using the Riemann-Lebesgue theorem, $\int_a^b f(k') e^{{\bf i}k'
\eta}\mbox{d}k'\rightarrow 0\ \mbox{as }\eta\rightarrow \infty \mbox{ provided that
 } \int_a^b |f(k')|\mbox{d}k'$ exists. Since $sin^2 \alpha=(1-cos 2\alpha)/2$ and 
$cos^2 \alpha=(1+cos 2\alpha)/2$, the integration yields a linear increase in time
for the pressure variance. This is consistent with the numerical solutions of the
linearized inviscid equations presented in Fig. 3. 

Similar to the pressure variance, it can be shown that for finite values of 
$\frac{1}{(M_{t_0}S)^2}$, most of the contribution to the dilatation variance 
and kinetic energy in $x_2$ direction, $K_2$, comes from the outer region. From the
solution obtained for the pressure mode, approximations for the asymptotic behavior
of the dilatation and $u_2$ velocity modes in the outer region can be derived using
equations

\begin{eqnarray}
\hat{\Delta}&=&-\frac{(M_{t_0}S)}{2 K_0} \hat{p}'_{,\eta}\\
\hat{u}_2&=&\frac{\mbox{\boldmath $i$}\hat{\Delta}_{,\eta}}{2k_1}-
 \frac{\mbox{\boldmath $i$}k^2\hat{p}'}{2<\rho>(M_{t_0}S)k_1}.
\end{eqnarray}

\noindent
The relations obtained are $\hat{\Delta} =\phi_{1\Delta}\ \hat{p}'|_{\eta=0}+
\phi_{2\Delta}\ \hat{p}_{,\eta}'|_{\eta=0}+\phi_{3\Delta}\ \hat{p}'_{,\eta\eta}|_{\eta=0}$ 
and $\hat{u}_2 =\phi_{1u2}\ \hat{p}'|_{\eta=0}+\phi_{2u2}\ \hat{p}_{,\eta}'|_{\eta=0}+
\phi_{3u2}\ \hat{p}'_{,\eta\eta}|_{\eta=0}$ with $\phi_{i\Delta}$ and $\phi_{iu2}$ given by

\begin{eqnarray}
\phi_{1\Delta}&\approx& (\zeta/2)^{3/4}\ cos(\zeta-\frac{3 \pi}{8})
  \Gamma\left(\frac{3}{4}\right)\left(\frac{k_1'^2 M_{t_0}^2S^2}{2<\rho>K_0^3}\right)^{1/4}
\label{lind1}\\
\phi_{2\Delta}&\approx& (\zeta/2)^{3/4}\ sin(\zeta-\frac{\pi}{8})
  \Gamma\left(\frac{3}{4}\right)\frac{M_{t_0}S}{\sqrt{2\pi}K_0}\\
\phi_{3\Delta}&\approx& (\zeta/2)^{3/4}\ sin(\zeta-\frac{3 \pi}{8})
  \Gamma\left(\frac{1}{4}\right)\frac{1}{4}\left(\frac{M_{t_0}^3S^3}{\pi k_1'}\right)^{1/2} 
  \left(\frac{<\rho>}{8K_0^5}\right)^{1/4}
\label{lind3}\\
\phi_{1u2}&\approx& \mbox{\boldmath $i$}(\zeta/2)^{1/4}\ cos(\zeta-\frac{3 \pi}{8})
  \Gamma\left(\frac{3}{4}\right)\frac{k_1'}{2<\rho>M_{t_0}S}
\label{linu21}\\
\phi_{2u2}&\approx& \mbox{\boldmath $i$} (\zeta/2)^{1/4}\ sin(\zeta-\frac{\pi}{8})
  \Gamma\left(\frac{3}{4}\right)\left(\frac{k_1'}{\pi M_{t_0}S}\right)^{1/2}
  \frac{1}{2K_0}
\label{linu22}\\
\phi_{3u2}&\approx& \mbox{\boldmath $i$} (\zeta/2)^{1/4}\ sin(\zeta-\frac{3 \pi}{8})
  \Gamma\left(\frac{1}{4}\right)\frac{1}{4}\left(\frac{1}{2\pi <\rho>K_0}\right)^{1/2}.
\label{linu23}
\end{eqnarray}

Again, using the Riemann-Lebesgue theorem, Eqs. (\ref{lind1})-(\ref{linu23}) can be integrated
over the wavenumber space to yield an $\eta^3$ increase in time for the dilatation variance,
$<\Delta'^2>$, and a linear increase for $<u_2^2>$. This represents an important difference
compared to the incompressible case, where it it is known that for large times the 
$u_2$ velocity variance decreases in time as $\ln(4\eta)/(4\eta)$.~\citep{Rogers91} It is shown 
below that the increase in time in the compressible case is due to the dilatational component.

In the incompressible case, the influence from the inner region to $<u_1^2>$ and 
$<u_3^2>$ is significant and at large times $<u_1^2>\approx 2 \ln 2\ \eta$ and 
$<u_3^2>\pi^2/8\ln\eta-C$.~\citep{Rogers91} For the compressible case, after using 
the approximation $k_2\approx-\eta k_1'$, the transport equations for $\hat{u}_1$ 
and $\hat{u}_3$ can be solved analytically and for large values of $\eta$ 

\begin{eqnarray}
\hat{u}_1&\sim&\hat{u}_1|_{\eta=0}+ \frac{d_{1}}{\sqrt{\eta}}\ 
  sin(\frac{k_1'\eta^2}{M_{t_0}S}\sqrt{\frac{K_0}{2<\rho>}}+d_{2})\\
\hat{u}_3&\sim&\hat{u}_3|_{\eta=0}+ \frac{e_{1}}{\sqrt{\eta}}\ 
  sin(\frac{k_1'\eta^2}{M_{t_0}S}\sqrt{\frac{K_0}{2<\rho>}}+e_{2})
\end{eqnarray}

\noindent
with the coefficients $d_i$ and $e_i$ depending on $(M_{t_0}S)$ and the wavenumber.
Therefore, in the outer region, for large times, $\hat{u}_1$ and $\hat{u}_3$ become
constant. Numerical solutions of the RDT equations indicate that the contributions 
from the inner region to $<u_1^2>$ and $<u_3^2>$ remain dominant for the 
compressible case and $K_1$ and $K_3$ increase in time, as shown in Fig. 4. However, 
as the compressibility parameter $M_{t_0}S$ increases, these contributions decrease
their importance resulting in a reduced rate of increase of the turbulent kinetic 
energy in $x_1$ and $x_3$ directions. This explains the ``stabilizing'' effect of 
compressibility on the evolution of the turbulent kinetic energy found
in the RDT study of Simone \etal.~\cite{SCC97} 
 
From the solution corresponding to the Fourier mode of the dilatation, the dilatational 
parts of the velocity modes can be obtained as 
$\hat{u}_{i_d}=-\frac{\mbox{\boldmath $i$}k_i}{k^2} \hat{\Delta}$.
Numerical solutions of the RDT equations indicate that the inner region has a negligible
contribution to the variances of the dilatational velocity components. In the outer region,
for large values of $\eta$, the time variation of the dilatational parts of the velocity 
modes can be derived from formulas \ref{lind1}-\ref{lind3} as

\begin{eqnarray}
\hat{u}_{1_d}&\sim& \frac{c_{11}}{\sqrt{\eta}}\ 
  sin(\frac{k_1'\eta^2}{M_{t_0}S}\sqrt{\frac{K_0}{2<\rho>}}+c_{12})
\label{linu1}\\
\hat{u}_{2_d}&\sim& c_{21}\sqrt{\eta}\ 
  sin(\frac{k_1'\eta^2}{M_{t_0}S}\sqrt{\frac{K_0}{2<\rho>}}+c_{22})
\label{linu2}\\
\hat{u}_{3_d}&\sim& \frac{c_{31}}{\sqrt{\eta}}\ 
  sin(\frac{k_1'\eta^2}{M_{t_0}S}\sqrt{\frac{K_0}{2<\rho>}}+c_{23}),
\label{linu3}
\end{eqnarray}

\noindent
where the constants $c_{ij}$ depend on $(M_{t_0}S)$ and the wavenumber. 

\section{Comparison with DNS}

Although the early time response of the flow to the presence of the shear is correctly
captured by the linearized equations, the long time evolution of the flow can become
very different than that obtained from the full nonlinear equations. However, there are
quantities which follow the RDT predictions even at long times. This is the case 
with various correlation coefficients (e.g. the Reynolds stress correlation coefficient) 
in the incompressible case.~\citep{Rogers91} For the compressible case, the effect of 
compressibility on the evolution of the turbulent kinetic energy is correctly captured by the 
linearized equations.~\citep{SCC97}  

It is shown below that other characteristics of the dilatational field are captured by the
linearized equations. In particular, it is shown that the amplification of the dilatational kinetic
energy in the direction of the shear and the anisotropy of the normal components of the 
dilatational dissipation rate tensor are captured by the RDT equations. Furthermore, 
DNS results show that the transverse even order derivative moments of the
velocity field in both compressible and incompressible cases are anisotropic and the anisotropy
increases with the order of the moment. Thus, the anisotropy of the higher order moments may persist at
large values of the Reynolds number, in contrast with the local isotropy principle.~\cite{MY75}
Moreover, the DNS results indicate that the odd order normalized $x_2$ derivative moments of $u_1$ approach  
a constant different than zero at large times. These results can also be explained using the RDT solutions.

\subsection{Dilatational kinetic energy and dissipation anisotropy}

By comparing formulas
\ref{linu1}-\ref{linu3} with solutions obtained for the velocity modes, it can be seen that at
long times $\hat{u}_1\approx\hat{u}_{1_s}$, $\hat{u}_2\approx\hat{u}_{2_d}$, and
$\hat{u}_3\approx\hat{u}_{3_s}$. It is obtained that most of the dilatational kinetic
energy resides in the direction of the mean velocity gradient,
consistent with the numerical solution of the linearized Eqs.
(\ref{colin})-(\ref{enlin}) (Fig. 5a). The addition of the viscous terms
decreases the dominance of $K_{2_d}$ component compared to the other
two components, as can be seen in Fig. 5(b). A further decrease of
the relative magnitude of $K_{2_d}$ is obtained in the fully nonlinear
case (Fig. 5c). Nevertheless, as Fig. 5(c) shows, the 
dilatational kinetic energy in the direction of the mean velocity gradient obtained
in DNS is larger than in the other two directions, in agreement with
the previous studies.~\citep{BMR91,FB97,LJM02} The above analysis indicates that 
the amplification of the dilatational kinetic energy in the direction of the mean
shear can be explained by linear effects. 

Previous numerical simulations of compressible homogeneous shear 
flow~\citep{BMR91,LM01,LJM02} indicate that the dilatational dissipation rate tensor 
behaves very differently than its solenoidal counterpart. Thus, the dilatational
dissipation rate is amplified in the direction of the mean shear. This behavior was 
previously attributed to the formation of shocklets aligned preferentially in $x_2$ 
direction. Livescu and Madnia~\cite{LM01} performed simulations at different values of
the turbulent Mach number and obtained no significant decrease in the anisotropy
of the dilatational dissipation rate tensor in low turbulent Mach number simulations.
Therefore, the amplification of the dilatational dissipation rate in the direction of 
the mean shear is not associated with the presence of shocklets. This conclusion is 
also obtained in Ref.~\onlinecite{BZ92}. Figure 6 shows that $\epsilon_{22_d}$ becomes more
important compared to $\epsilon_{11_d}$ and $\epsilon_{33_d}$ as $M_{t_0}S$, which
is proportional to the distortion Mach number, increases. After multiplying the 
solutions for the dilatational velocity modes (Eqs. \ref{linu1}-\ref{linu3}) by 
$\mbox{\boldmath $i$}k_i$ and performing the integration over the wavenumber space, it
can be seen that in the RDT limit $<\frac{\partial u_{2_d}}{\partial x_2}
\frac{\partial u_{2_d}}{\partial x_2}>$ becomes much larger than $<\frac{\partial 
u_{1_d}}{\partial x_1}\frac{\partial u_{1_d}}{\partial x_1}>$ and $<\frac{\partial 
u_{3_d}}{\partial x_3}\frac{\partial u_{3_d}}{\partial x_3}>$. Numerical solutions of 
the RDT equations in both inviscid and viscous cases confirm this result. Therefore, 
the anisotropy of the normal components of the dilatational dissipation rate tensor 
can be associated with a linear mechanism.

\subsection{Higher order derivative moments}

In the outer region $k_2$ increases continuously, while it is very small in the inner region
(where $k_1'\eta\approx k_2'$). Using the solutions provided in the previous section
for the solenoidal and dilatational components of the velocities, it can
be shown that after the multiplication by $k_2$, the contribution from the outer region
becomes dominant at long times. Therefore, the $x_2$ derivatives of the velocity components
become much larger than the derivatives in the other two directions. As a result, all the even order 
derivative statistics involving $x_2$ derivatives become amplified and this effect is 
more pronounced for higher order statistics. Figure 7 compares moments of the transverse 
derivatives, $\left<\left(\frac{\partial u_{1_d}}{\partial x_j}\right)^n\right>$ where
$j=2,3$ and $n=2$, $4$ and $6$, of the dilatational velocity in $x_1$ direction. It 
can be seen that, as the distortion Mach number increases and the RDT limit is 
approached, the moments containing $x_2$ derivatives become much larger than the 
corresponding $x_3$ derivative statistics. Moreover, in agreement with the RDT 
predictions, this anisotropy is strongly amplified for higher order moments. This
again suggests that the small scale anisotropy of the dilatational motions found in 
the DNS results is produced by a linear mechanism. On the other hand, the results presented show 
that the higher order transverse moments are increasingly more anisotropic as the order of the moment 
increases. Thus, even if the anisotropy decreases with Reynolds number, some degree of anisotropy 
will persist for higher order transverse derivative moments at large values of the Reynolds number.

Although the inner region has a significant contribution to the one-point statistics
of the solenoidal velocity components in $x_1$ and $x_3$ directions, this contribution becomes 
negligible for the statistics involving $x_2$ derivatives. As explained above, 
in the outer region $\hat{u}_{1_s}$ and $\hat{u}_{3_s}$ become constant at large times so that
after the multiplication with $k_2$ they increase linearly. Therefore, the outer region yields
a quadratic increase in time for $<\frac{\partial u_{1_s}}{\partial x_2}\frac{\partial u_{1_s}}{\partial x_2}>$ 
and $<\frac{\partial u_{3_s}}{\partial x_2}\frac{\partial u_{3_s}}{\partial x_2}>$. The contribution from the
inner region does not increase after the multiplication with $k_2$ and the inner region yields
a less than quadratic increase for the above quantities. Similar results are obtained for higher
order moments. Thus, similar to the dilatational velocities, in the RDT limit the even order transverse 
derivatives of the solenoidal velocities in $x_1$ and $x_3$ directions are anisotropic and become 
even more anisotropic as the order of the statistics increases. Figure 8 confirms that this 
effect is also present in the DNS results for the derivative moments of $u_{1_s}$ and similar results
are obtained for the derivative moments of $u_{3_s}$. 

A comparison between Figs. 7 and 8 indicates that 
the transverse solenoidal derivative moments are less anisotropic than their dilatational counterparts.
Again, it can be shown that this is also in agreement with the RDT predictions. Thus, since 
$\hat{u}_{1_s}$ becomes constant in the outer region, it yields that $k_2\hat{u}_{1_s}\sim\eta$
and $\left<\left(\frac{\partial u_{1_s}}{\partial x_2}\right)^2\right>\sim \eta^2$. However,
the $x_3$ derivative retains influence mostly from the inner region and 
$\left<\left(\frac{\partial u_{1_s}}{\partial x_3}\right)^2\right>\sim \eta$. On the other
hand $\hat{u}_{1_d}$ retains influence mostly from the outer region. The Riemann-Lebesgue
theorem can be used to integrate formula \ref{linu1} multiplied by $k_2$ and $k_3$ to obtain
for the dilatational component that 
$\left<\left(\frac{\partial u_{1_d}}{\partial x_2}\right)^2\right>\sim \eta$ and
$\left<\left(\frac{\partial u_{1_d}}{\partial x_3}\right)^2\right>\sim 1/\eta$. Similar results
are obtained for higher order statistics. Therefore, in the RDT limit, the even order transverse
derivatives of the $u_{1d}$ are more anisotropic than their solenoidal counterparts for large but
finite values of $\eta$ (Fig. 9). Similar results are obtained for the velocity in $x_3$
direction.

Using the same analysis as above, it can be shown that the odd order moments of the
$x_2$ derivatives of $u_{1_s}$ also have most of the contributions from the outer 
regions. Thus, after the multiplication with $k_2^n\approx (-\eta k_1')^n$, the solution
in the outer region becomes much larger than that obtained in the inner region as $\eta
\rightarrow \infty$. Note that $|u_{1_s}|$ increases in time as $\eta^{1/2}$ in the inner
region and the multiplication by $k_2$ does not increase this rate. Since in the outer 
region $\hat{u}_{1_s}$ becomes constant at large times, it yields that the moment
of order $n$ increases in time as $\eta^n$, and, therefore, the normalized odd order 
moments, $\left<\left(\frac{\partial u_{1_s}}{\partial x_2}\right)^n\right>/
\left[\left<\left(\frac{\partial u_{1_s}}{\partial x_2}\right)^2\right>\right]^
{n/2}$, with $n$ odd, should approach a constant different than zero at large 
times. As Fig. 10 shows, this prediction is in general agreement with the DNS data. 
The same analysis can not be used for $u_{3_s}$. It was shown that in the outer
region $\hat{u}_{3_s}$ approaches its initial value. Therefore, after multiplication
by $k_2^n\approx (-\eta k_1')^n$ the odd order moments of the $x_2$ derivative, would 
approach values proportional to those of the initial odd order moments of the $x_1$
derivative. However, for isotropic initial conditions, these values are zero. This
is consistent with the overall symmetry of the problem, which requires that the odd
order $x_1$ and $x_2$ moments of the $u_3$ velocity remain zero at all times.

For the incompressible case, complete analytical solutions are known for the velocity 
field.~\citep{Townsend76,Rogers91} Although most of the contributions to the velocity 
variances in $x_1$ and $x_3$ directions come from the inner region, it is easy to show that
the statistics involving $x_2$ derivatives have contributions mostly from the outer region,
where at large times $k_2\approx -\eta k_1'$. Moreover, the behavior of the velocity modes 
in the outer region is similar to that obtained above for the solenoidal velocity modes. 
Consequently, in the RDT limit the transverse even order moments of the velocity 
derivatives are anisotropic, and the anisotropy increases for higher order moments. Similar
to the solenoidal moments for the compressible case and consistent with the RDT 
predictions, the incompressible DNS results presented in Fig. 11 show an increase in the 
anisotropy for higher order transverse derivatives moments of $u_1$. Similar results are 
obtained for $u_3$. Therefore, it is expected that for higher order moments the anisotropy 
will persist at large Reynolds numbers. 

In addition, similar to the solenoidal velocity for the compressible case, the RDT 
solutions for the incompressible case predict that the odd order normalized $x_2$ 
derivative moments of $u_1$ and $u_3$ should approach a constant different than zero at 
large times. The DNS results are similar to those shown in Fig. 10 for the solenoidal 
velocity field in the compressible case and support the RDT prediction. As explained above,
as the Reynolds number increases the small scales become more energetic and the flow 
departs from the RDT limit (see also Fig. 1). For very large 
values of the Reynolds number, the postulate of local isotropy requires that the odd order transverse derivative 
moments should approach zero. However, since the linear effects lead to a $\eta^n$ increase in time for the 
$x_2$ derivative moments, for higher order moments the linear 
contribution may become dominant at long times even at large values of the Reynolds number. Therefore, the 
higher order normalized transverse $x_2$ derivative moments could have persistent non-zero values at large 
Reynolds numbers. The present numerical simulations do not have a large enough Reynolds number to examine 
this hypothesis, however the analysis presented is consistent with the experimental results of 
Shen and Warhaft.~\cite{SW00} At high Reynolds numbers, the experimental results indicate that lower odd order 
moments of $\frac{\partial u_1}{\partial x_2}$ become small as the Reynolds number increases. However, 
this tendency is not observed for higher order moments. As suggested by Schumacher \etal,~\cite{SRY03}
the small-scale intermittency might play a role on the observed behavior of the higher order moments.
Nevertheless, the persistent anisotropy of the higher odd order normalized derivative moments found
experimentally is in agreement with the present linear analysis. 

\section{Summary and Conclusions}

The structure of homogeneous turbulent shear flow is studied using data 
generated by DNS and an RDT analysis for both compressible and incompressible cases.
For the compressible case, simulations with different initial values of the distortion Mach number, $M_d$,
and Reynolds number are considered. Incompressible simulations are performed with the same 
initial pressure and velocity fields as for the compressible simulations.

Previous DNS studies~\citep{BMR91,LM01,LJM02} indicate that there is an anisotropy
among the dilatational kinetic energy components. Moreover, the results of
Livescu \etal~\cite{LJM02} suggest that the energy transfer through the nonlinear terms in the 
transport equations for $K_{i_d}$ is not responsible for this behavior. In order 
to verify that the anisotropy of the dilatational kinetic energy components can be explained by 
linear effects, the linearized equations are considered. The RDT equations are 
solved numerically, for both the inviscid and viscous cases, for a large range of
distortion Mach numbers and different types of initial conditions. The RDT
results are consistent with the DNS findings for the development of the 
dilatational field. 

Furthermore, for large times, analytical solutions are found for the
inviscid linearized equations in Fourier space. The integration of
these solutions over the wavenumber space can predict the behavior of
various statistics in real space. It is shown that the solutions
for the pressure, dilatation and dilatational velocities Fourier modes for finite
values of $k_1'$ oscillate in time and the amplitudes of oscillations increase.
 The analytical relations indicate that the magnitude of the Fourier mode of the 
dilatational velocity in the direction of the mean shear becomes much larger than
in the other two directions. 

It is known that for incompressible homogeneous turbulent shear flow in the RDT limit
most of the contributions to the velocity variances come from a narrow region in
the wavenumber space situated near $k_1'=0$.~\cite{Rogers91} A similar trend is
found for the pressure variance. However, it is shown that the contribution from this
region to the statistics of the velocity field involving $x_2$ derivatives for both 
compressible and incompressible cases and to all the statistics of the dilatational 
velocities becomes negligible at long times. For finite values of $k_1'$, the wave number in the 
direction of the mean shear increases continuously in time. Therefore, the derivatives in this direction 
become much larger than in the other two directions. As a result, in the
RDT limit the even order transverse derivative moments of both the dilatational
and solenoidal velocity fields for the compressible case and the even order transverse derivative 
moments of the velocity field for the incompressible case are anisotropic. Moreover, the anisotropy 
becomes amplified for higher order moments. The RDT analytical solutions also predict that the small 
dilatational scales are more anisotropic than their solenoidal counterparts, in agreement with the 
DNS results. Since the $x_2$ derivatives of the velocity field retain influence mostly from the region 
where the wavenumber in $x_2$ direction increases in time, it is shown that in the RDT limit
the odd order normalized $x_2$ derivative moments of $u_1$ for the incompressible case and 
$u_{1_s}$ for the compressible case should approach a constant different than zero 
at large times. This prediction is consistent with the DNS results.

As the Reynolds number increases the small scales become more energetic and for constant
mean distortion the flow departs the RDT limit. Therefore, it is expected that,
consistent with the postulate of local isotropy, the anisotropy of the small scales should 
decrease at large values of the Reynolds number, i.e. the odd order normalized $x_2$ derivative 
moments approach zero and the even order transverse moments become isotropic.
However, it is argued that, since in the RDT limit the odd order moments increase in time as
$\eta^n$, where $n$ is the order of the moment, for higher order moments the linear contribution
may be important even at large values of the Reynolds number. Thus, the RDT analysis suggests 
a persistent anisotropy of the higher odd order normalized derivative moments.
In addition, since the anisotropy of the even order transverse derivative moments increases with the 
order of the moment, it is expected that persistent anisotropy levels in the transverse higher 
even order derivative moments will be found at large values of the Reynolds number. Although high 
Reynolds number results will not be available from DNS in the near future, they are within the reach
of experiments and these predictions can be verified using experimental data.

\begin{acknowledgments}

This work was sponsored by the Donors of the Petroleum Research Funds
administrated by the American Chemical Society under Grant No. 35064-AC9
and the US Department of Energy.
Computational resources were provided by the National Center for Supercomputer 
Applications at the University of Illinois 
Urbana-Champaign, the Center for Computational Research at State University
of New York at Buffalo, and through the Institutional Computing Project,
Los Alamos National Laboratory.
\end{acknowledgments}

\newpage
\mbox{ }
\vskip 2cm
\begin{table}[h]
\begin{center}
\label{tab1}
\begin{tabular}{||c|c|c|c|c||}
\hline\noalign{\smallskip}
Case      & $Re_{\lambda_0}$ & $S_0^*$& $M_{d_0}^{\dagger}$\\
\noalign{\smallskip}\hline\noalign{\smallskip}
 1     & 21  &  4.86   & 1.46 \\ 
 2     & 21  &  7.29   & 2.19 \\ 
 3     & 21  &  9.71   & 2.91 \\ 
 4     & 21  &  14.6   & 4.37 \\ 
 5     & 21  &  21.9   & 6.56 \\ 
 6     & 50  &  17.4   & 5.22 \\ 
\noalign{\smallskip}\hline
\end{tabular}
\caption{Parameters for the DNS cases. $^{\dagger}$ Values corresponding to the compressible cases.
All compressible cases have $M_{t_0}=0.3$.}
\end{center}
\end{table}

\newpage
\mbox{ }
\vskip 2cm
\begin{table}[h]
\begin{center}
\label{tablin1}
\begin{tabular}{||c|c|c||}
\hline\noalign{\smallskip}
Type of initial conditions & $\hat{p}|_{\eta=0}$ & $\chi_0$  \\
\noalign{\smallskip}\hline\noalign{\smallskip}
L1   & Poisson equation & 0  \\ 
L2   &  0 & 0  \\ 
L3   & Poisson equation & 0.1  \\ 
L4   & Acoustic equilibrium & 0.1  \\ 
\noalign{\smallskip}\hline
\end{tabular}
\caption{Initial conditions for the linearized equations.}
\end{center}
\end{table}

\newpage
\section*{Figure captions}

{\bf Fig. 1 } Time evolution of $S^*_1/Re_{\lambda_1}$. 

{\bf Fig. 2.} Compressibility effects on the $k_1'$ variation of the pressure mode. 
Inviscid case, $\eta=10$, $k_2=1$, $k_3=0$, with initial conditions $\hat{p}|_{\eta=0}=1$,
$\hat{p}_{,\eta}|_{\eta=0}=0$, and $\hat{p}_{,\eta\eta}|_{\eta=0}=0$.
(a) Incompressible, (b) compressible, $M_{t_0}S=0.01$, and (c) compressible, $M_{t_0}S=1$.

{\bf Fig. 3.} Numerical solution of the inviscid RDT equations for the pressure
variance for $M_{t_0}S=4.6$ (thin lines) and $M_{t_0}S=31$ (thick lines) and 
different initial conditions. For clarity the results obtained for 
$M_{t_0}S=4.6$ are magnified 5 times.

{\bf Fig. 4.} Compressibility effect on the kinetic energy in $x_1$ and $x_3$ directions. 
Inviscid case, with L1 initial conditions.

{\bf Fig. 5.} Time evolution of the normalized dilatational kinetic energy
components for $M_{t_0}S=1.53,\ 3.1,$ and $4.6$. (a) Inviscid RDT, 
(b) viscous RDT, and (c) DNS. The curves correspond to cases C2, C4, and C5.

{\bf Fig. 6.} Time variation of the diagonal components of the dilatational 
dissipation rate tensor for $M_{t_0}S=1.53,\ 3.1,$ and $4.6$, corresponding to cases
C2, C4, and C5.

{\bf Fig. 7.} Even order transverse derivative moments of the dilatational velocity in
$x_1$ direction, $M_{ny}^d=\frac{\left<\left(\frac{\partial u_{1_d}}{\partial x_2}
\right)^n\right>}{\left<\left(\frac{\partial u_{1_d}}{\partial x_2}\right)^n\right>+
\left<\left(\frac{\partial u_{1_d}}{\partial x_3}\right)^n\right>}$ and 
$M_{nz}^d=\frac{\left<\left(\frac{\partial u_{1_d}}{\partial x_3}
\right)^n\right>}{\left<\left(\frac{\partial u_{1_d}}{\partial x_2}\right)^n\right>+
\left<\left(\frac{\partial u_{1_d}}{\partial x_3}\right)^n\right>}$.
(a) n=2, (b) n=4, and (c) n=6.

{\bf Fig. 8.} Even order transverse derivative moments of the solenoidal velocity in
$x_1$ direction, $M_{ny}^s=\frac{\left<\left(\frac{\partial u_{1_s}}{\partial x_2}
\right)^n\right>}{\left<\left(\frac{\partial u_{1_s}}{\partial x_2}\right)^n\right>+
\left<\left(\frac{\partial u_{1_s}}{\partial x_3}\right)^n\right>}$.
n=2 (no symbols), n=4 (open symbols), and n=6 (closed symbols).

{\bf Fig. 9.} RDT results for the second order transverse derivatives of $u_{1_s}$ and
$u_{1_d}$. L1 initial conditions, $M_{t_0}S=1.53$.

{\bf Fig. 10.} Odd order normalized transverse derivative moments of the solenoidal velocity in
$x_1$ direction, $S_{ny}=\frac{\left<\left(\frac{\partial u_{1_s}}{\partial x_2}
\right)^n\right>}{\left(\left<\left(\frac{\partial u_{1_s}}{\partial x_2}\right)^2\right>\right)^{n/2}}$.
a) n=3 b) n=5.

{\bf Fig. 11.} Even order transverse derivative moments of the velocity in
$x_1$ direction, incompressible case. $M_{ny}=\frac{\left<\left(\frac{\partial u_1}{\partial x_2}
\right)^n\right>}{\left<\left(\frac{\partial u_1}{\partial x_2}\right)^n\right>+
\left<\left(\frac{\partial u_1}{\partial x_3}\right)^n\right>}$ and 
$M_{nz}=\frac{\left<\left(\frac{\partial u_1}{\partial x_3}
\right)^n\right>}{\left<\left(\frac{\partial u_1}{\partial x_2}\right)^n\right>+
\left<\left(\frac{\partial u_1}{\partial x_3}\right)^n\right>}$.
(a) n=2, (b) n=4, and (c) n=6.

\newpage
\mbox{\bf Fig. 1}
\vskip 1cm
\begin{figure}[h]
\centering
\vskip 6cm
\includegraphics[height=8.5cm,clip]{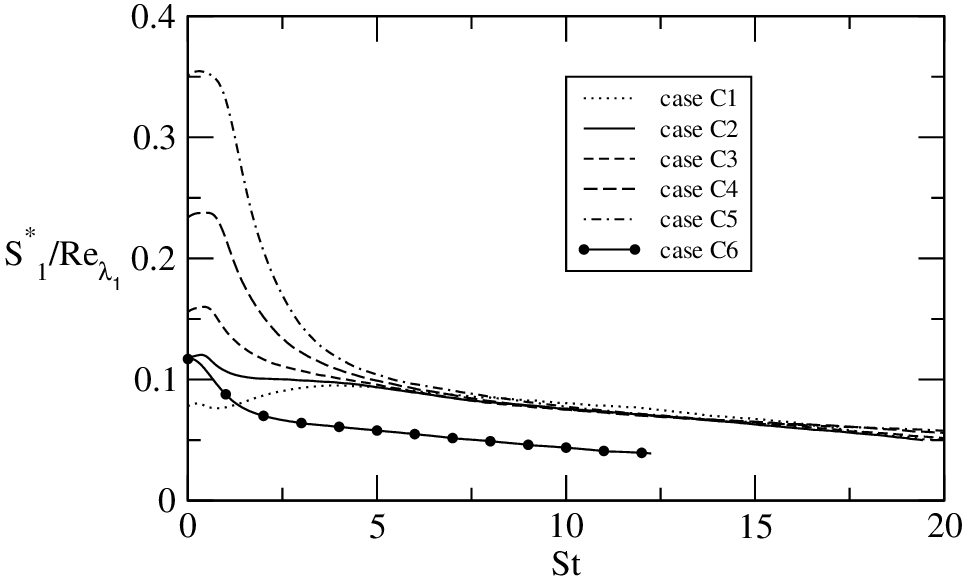}
\label{fig01}
\end{figure}

\newpage
\mbox{\bf Fig. 2}
\vskip 1cm
\begin{figure}[h]
\centering
\vskip 2cm
\includegraphics[height=17.5cm,clip]{fig1.eps}
\label{fig1}
\end{figure}

\newpage
\mbox{\bf Fig. 3}
\vskip 1cm
\begin{figure}[h]
\centering
\vskip 6cm
\includegraphics[height=8.5cm,clip]{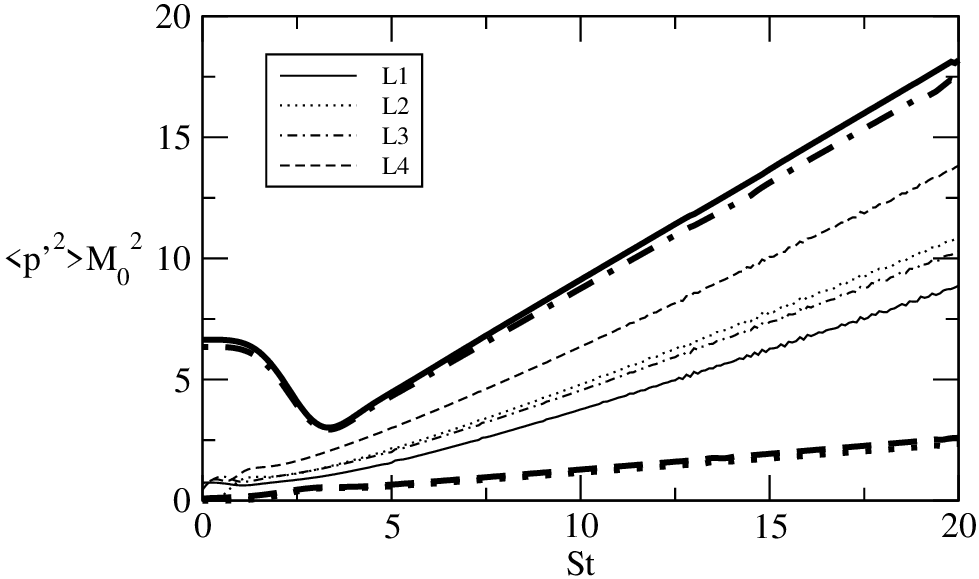}
\label{fig2}
\end{figure}

\newpage
\mbox{\bf Fig. 4}
\vskip 1cm
\begin{figure}[h]
\centering
\vskip 6cm
\includegraphics[height=8.5cm,clip]{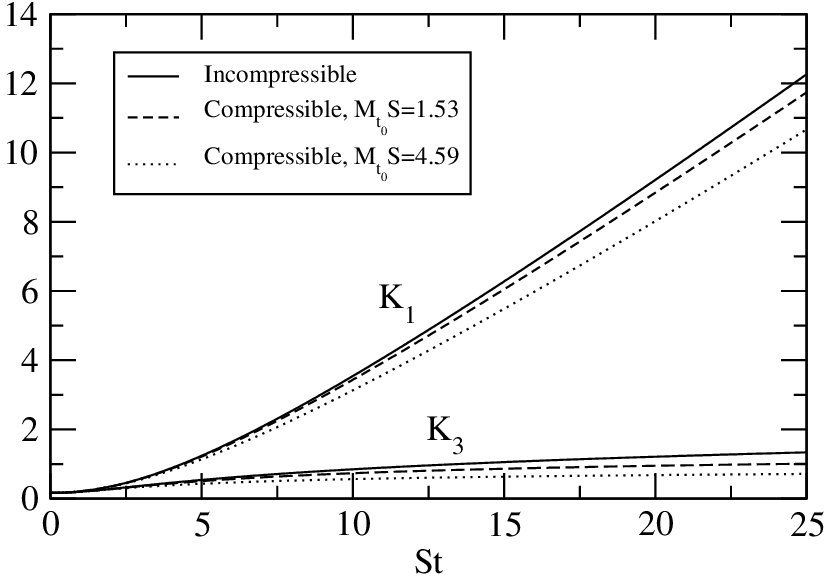}
\label{fig3}
\end{figure}

\newpage
\mbox{\bf Fig. 5}
\vskip 1cm
\begin{figure}[h]
\centering
\vskip 2cm
\includegraphics[height=17.5cm,clip]{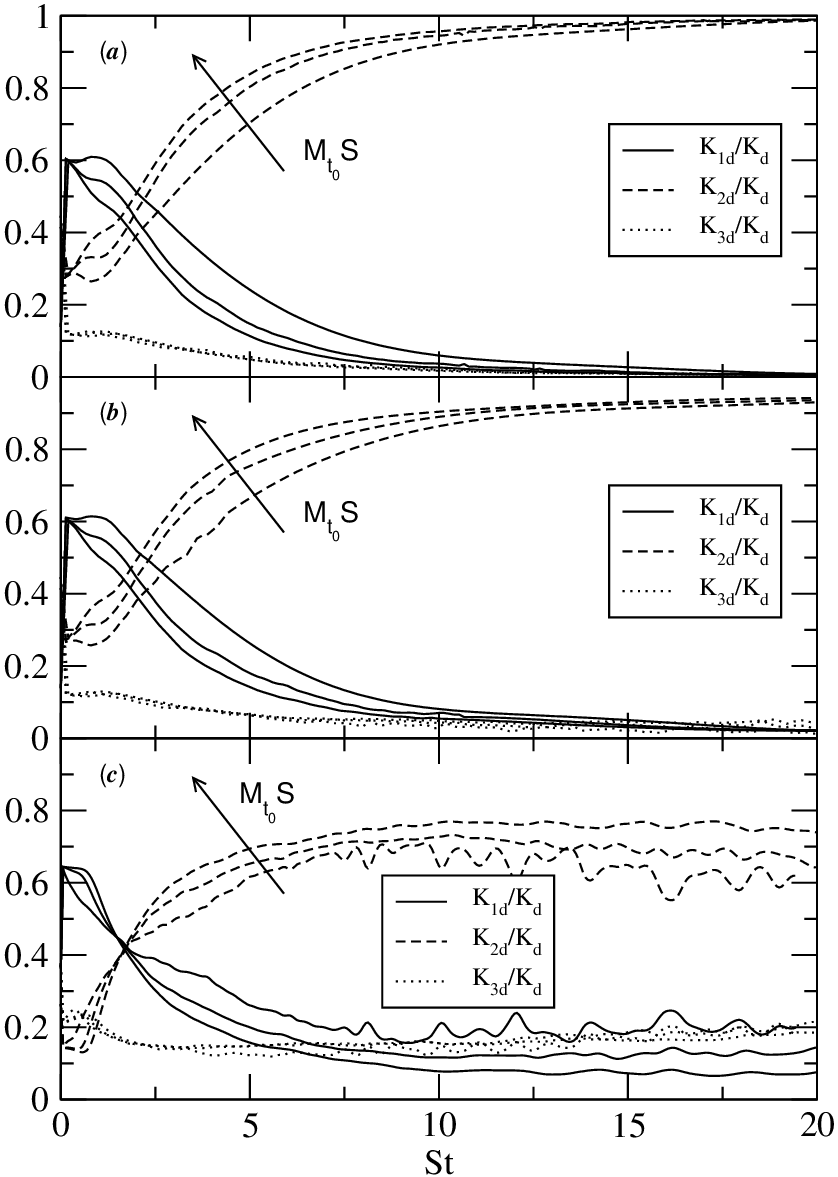}
\label{fig4}
\end{figure}

\newpage
\mbox{\bf Fig. 6}
\vskip 1cm
\begin{figure}[h]
\centering
\vskip 6cm
\includegraphics[height=8.5cm]{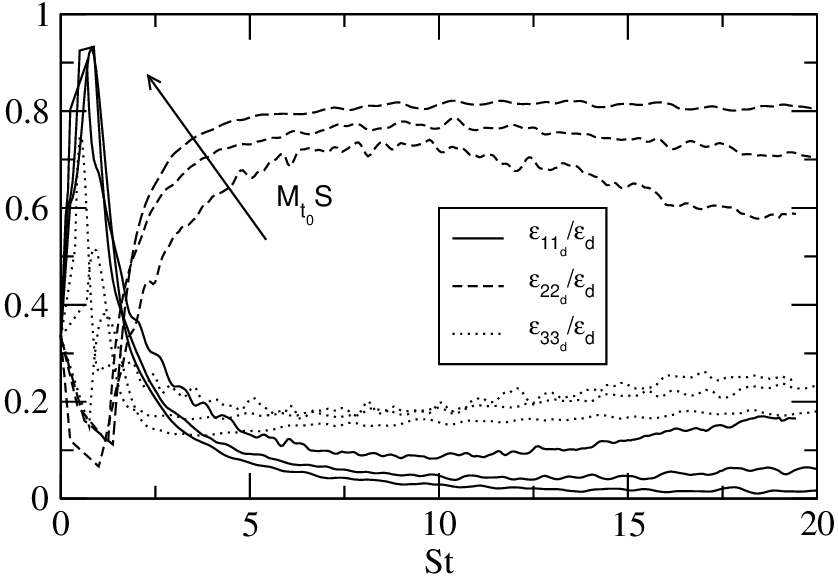}
\label{fig5}
\end{figure}

\newpage
\mbox{\bf Fig. 7}
\vskip 1cm
\begin{figure}[h]
\centering
\vskip 2cm
\includegraphics[height=17.5cm,clip]{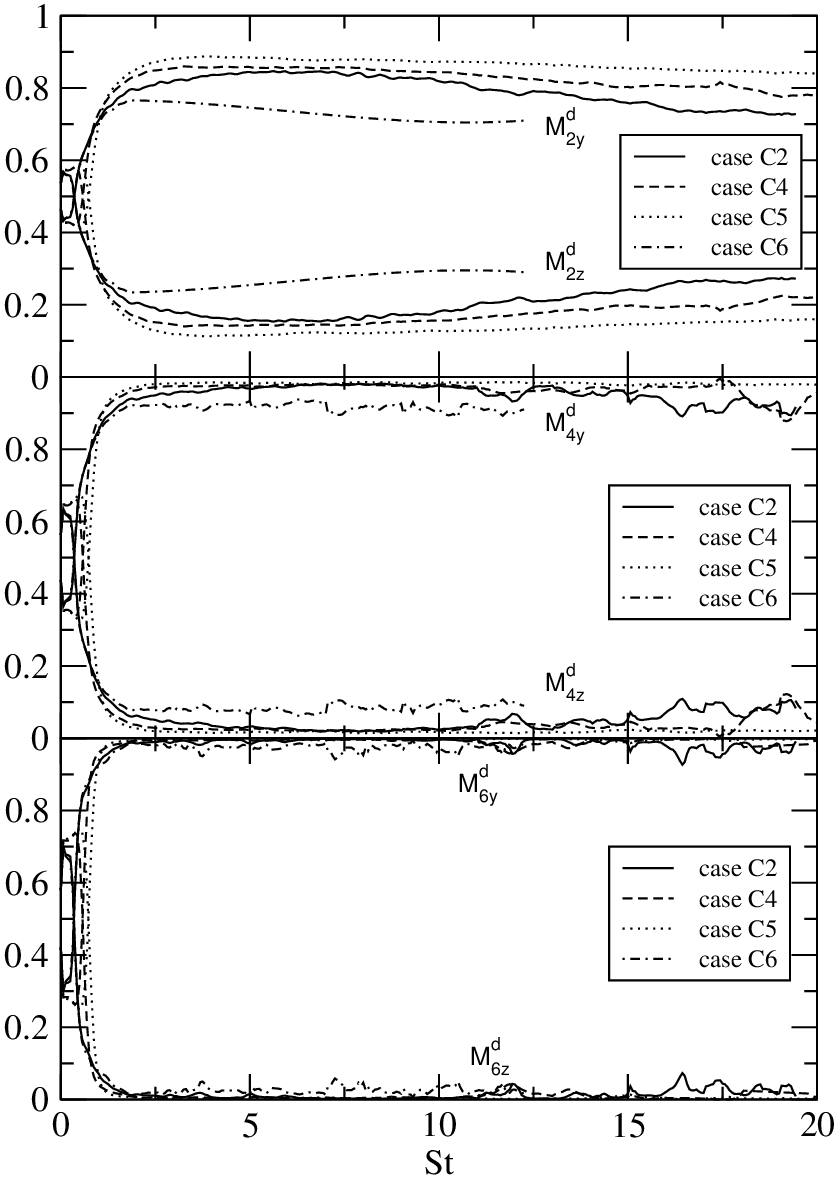}
\label{fig6}
\end{figure}

\newpage
\mbox{\bf Fig. 8}
\vskip 1cm
\begin{figure}[h]
\centering
\vskip 6cm
\includegraphics[height=8.5cm,clip]{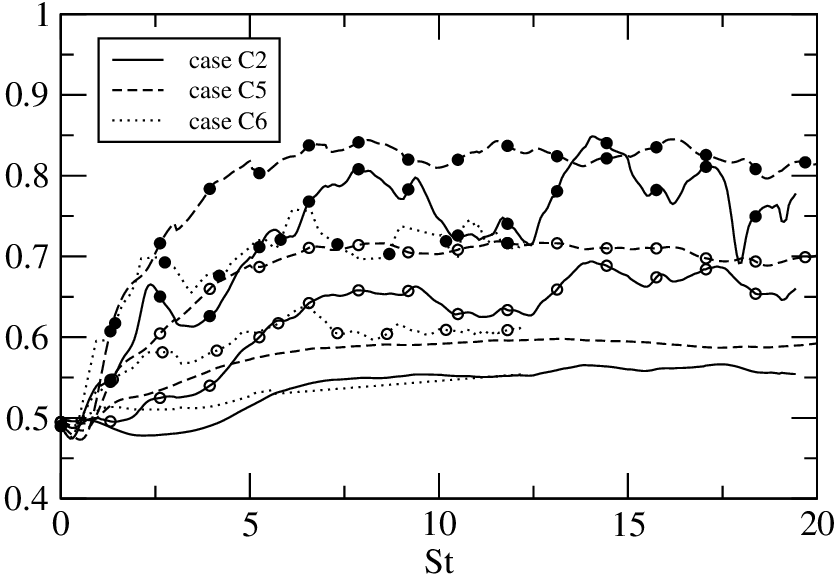}
\label{fig7}
\end{figure}

\newpage
\mbox{\bf Fig. 9}
\vskip 1cm
\begin{figure}[h]
\centering
\vskip 6cm
\includegraphics[height=8.5cm,clip]{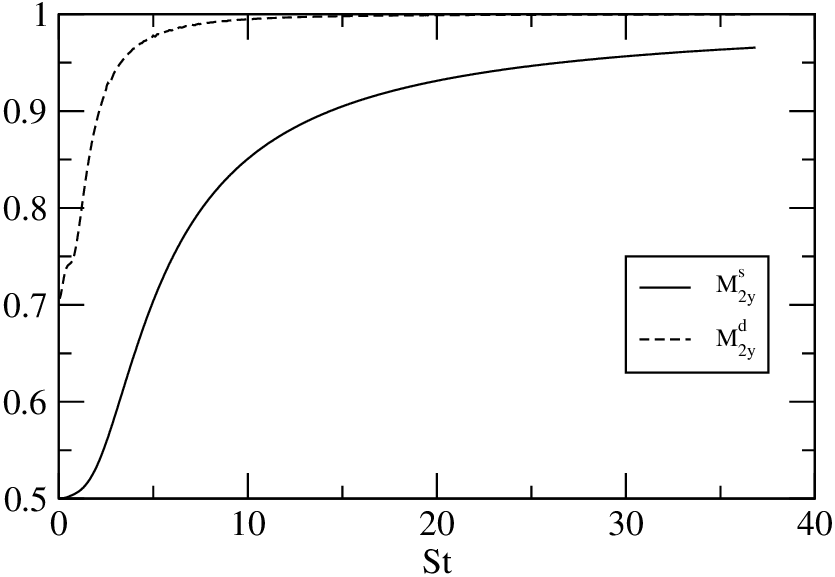}
\label{fig8}
\end{figure}

\newpage
\mbox{\bf Fig. 10}
\vskip 1cm
\begin{figure}[h]
\centering
\vskip 2cm
\includegraphics[height=17.5cm,clip]{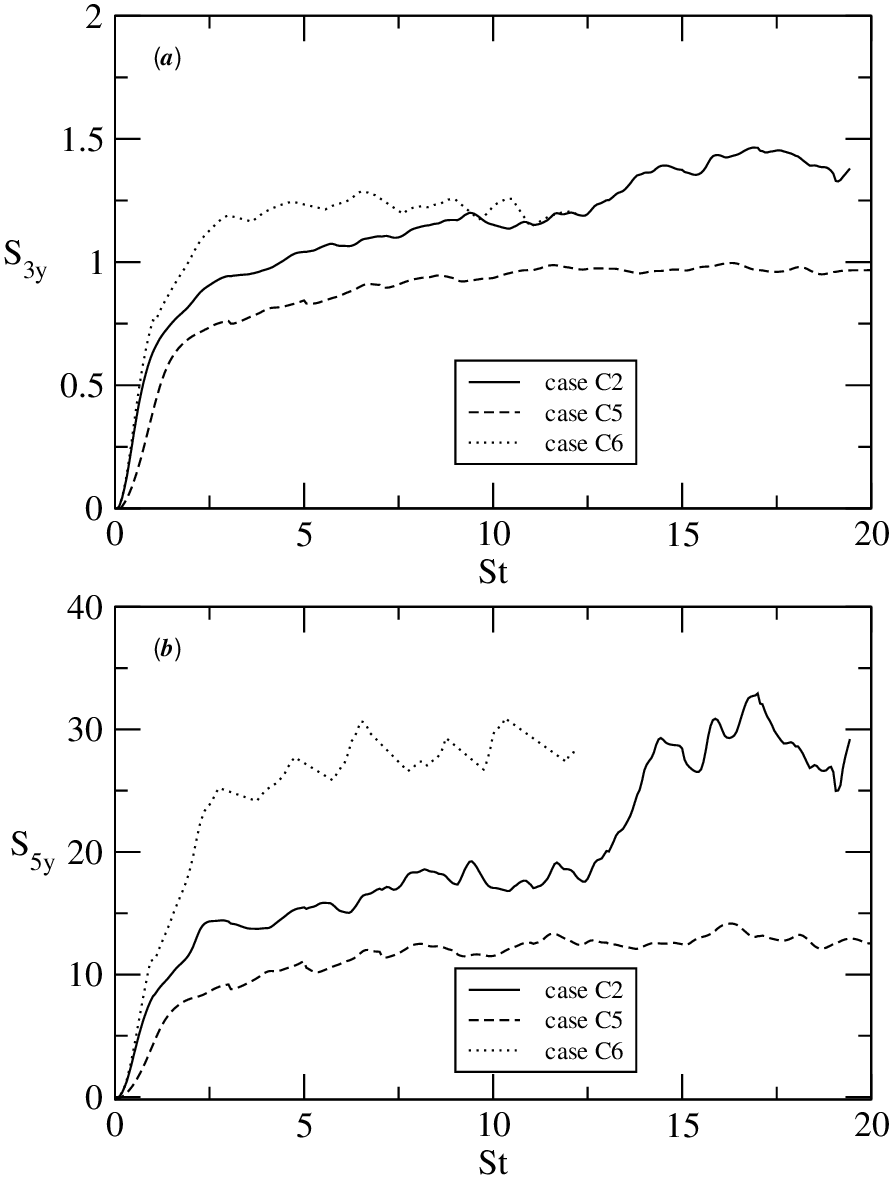}
\label{fig9}
\end{figure}

\newpage
\mbox{\bf Fig. 11}
\vskip 1cm
\begin{figure}[h]
\centering
\vskip 2cm
\includegraphics[height=17.5cm,clip]{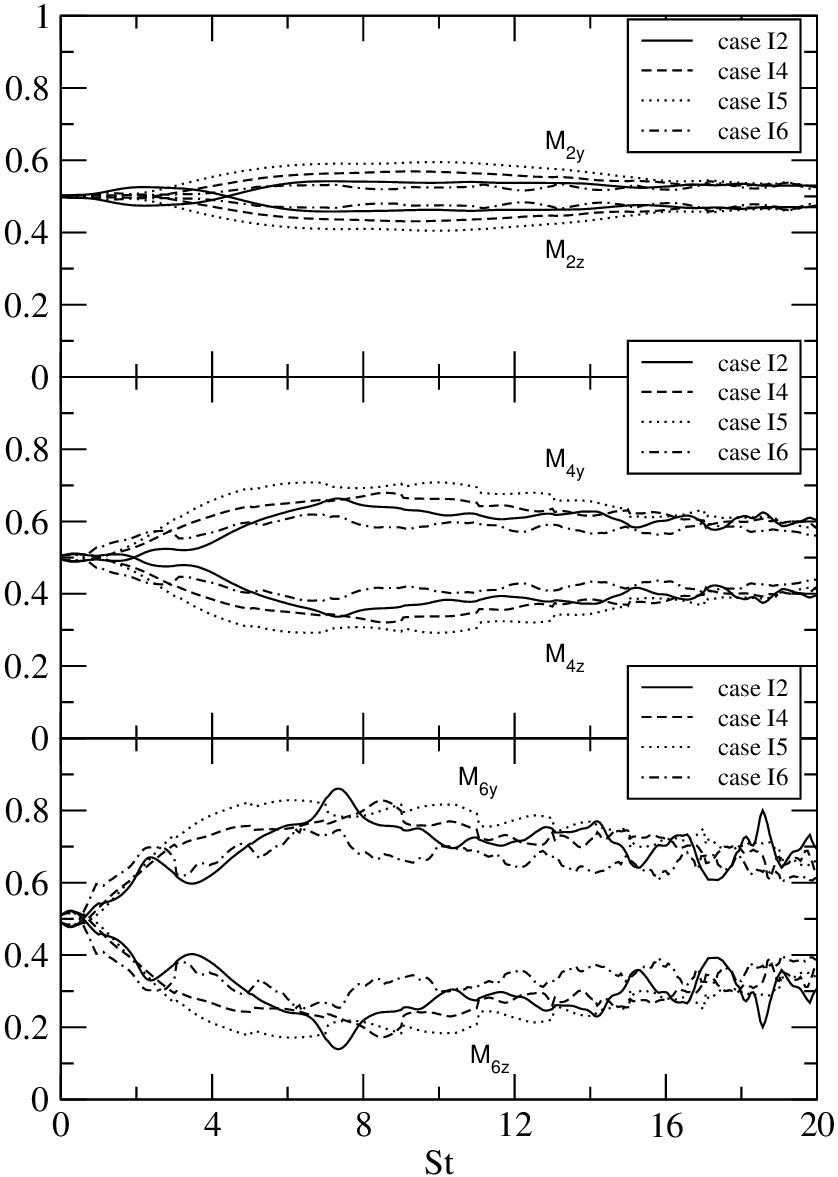}
\label{fig10}
\end{figure}

\end{document}